\documentclass[conference]{IEEEtran}
\IEEEoverridecommandlockouts
\usepackage{cite}
\usepackage{amsmath,amssymb,amsfonts}
\usepackage{graphicx}
\usepackage{textcomp}
\usepackage{xcolor}
\usepackage{url}

\ifCLASSOPTIONcompsoc
 \usepackage[caption=false,font=normalsize,labelfont=sf,textfont=sf]{subfig}
\else
 \usepackage[caption=false,font=footnotesize]{subfig}
\fi

\usepackage{algpseudocode}

\usepackage{acronym}
\usepackage{blindtext}
\usepackage{multicol,lipsum}

\newacro{vt}[VT]{Virtual Triggering}
\newacro{sca}[SCA]{Side-Channel Attack}
\newacro{scca}[SCCA]{Side-Channel Cryptographic Attack}
\newacro{snr}[SNR]{Signal to Noise Ratio}
\newacro{poi}[POI]{Point of Interest}
\newacroplural{poi}[POIs]{Points of Interest}
\newacro{ml}[ML]{Machine Learning}
\newacro{dl}[DL]{Deep Learning}
\newacro{rf}[RF]{Radio Frequency}
\newacro{aes}[AES]{Advanced Encryption Standard}
\newacro{sdr}[SDR]{Software Defined Radio}
\newacro{co}[CO]{Cryptographic Operation}
\newacro{cp}[CP]{Cryptographic Process}
\newacroplural{cp}[CPs]{Cryptographic Processes}
\newacro{pge}[PGE]{Partial Guessing Entropy}
\newacro{dpa}[DPA]{Differential Power Analysis}
\newacro{cpa}[CPA]{Correlation Power Analysis}
\newacro{mia}[MIA]{Mutual Information Analysis}
\newacro{em}[EM]{Electromagnetic}
\newacro{fc}[FC]{Frequency Component}
\newacro{da}[DA]{Differential Analysis}
\newacro{ca}[CA]{Correlation Analysis}
\newacro{puf}[PUF]{Physical Unclonable Function}

\def\BibTeX{{\rm B\kern-.05em{\sc i\kern-.025em b}\kern-.08em
    T\kern-.1667em\lower.7ex\hbox{E}\kern-.125emX}}

\title{Virtual Triggering: a Technique to Segment Cryptographic Processes in Side-Channel Traces}

\makeatletter
\newcommand{\linebreakand}{%
  \end{@IEEEauthorhalign}
  \hfill\mbox{}\par
  \mbox{}\hfill\begin{@IEEEauthorhalign}
}
\makeatother

\author{
    \IEEEauthorblockN{1\textsuperscript{st} Jeremy Guillaume}
    \IEEEauthorblockA{\textit{IETR UMR CNRS 6164} \\
    \textit{CentraleSupélec Rennes Campus}\\
    35576 Cesson-Sevigné, France \\
    jeremy.guillaume@centralesupelec.fr}
    \and
    \IEEEauthorblockN{2\textsuperscript{nd} Maxime Pelcat}
    \IEEEauthorblockA{\textit{IETR UMR CNRS 6164} \\
    \textit{INSA Rennes}\\
    35700 Rennes, France \\
    maxime.pelcat@insa-rennes.fr}
    \linebreakand 
    \IEEEauthorblockN{3\textsuperscript{rd} Amor Nafkha}
    \IEEEauthorblockA{\textit{IETR UMR CNRS 6164} \\
    \textit{CentraleSupélec Rennes Campus}\\
    35576 Cesson-Sevigné, France \\
    amor.nafkha@centralesupelec.fr}
    \and
    \IEEEauthorblockN{4\textsuperscript{th} Rubén Salvador}
    \IEEEauthorblockA{\textit{IETR UMR CNRS 6164} \\
    \textit{CentraleSupélec Rennes Campus}\\
    35576 Cesson-Sevigné, France \\
    ruben.salvador@centralesupelec.fr}
}

\begin{document}

\IEEEoverridecommandlockouts
\IEEEpubid{\makebox[\columnwidth]{978-1-6654-8524-1/22/\$31.00~\copyright2022 IEEE \hfill} \hspace{\columnsep}\makebox[\columnwidth]{ }}

\maketitle

\IEEEpubidadjcol

\begin{abstract}
\acfp{sca} exploit data correlation in signals leaked from devices to jeopardize confidentiality. 
Locating and synchronizing segments of interest 
in traces from \acp{cp} is a key step of the attack.
The most common method consists in generating a trigger signal to indicate to the attacker the start of a \ac{cp}. 
This paper proposes a method called \ac{vt} that
removes the need for the trigger signal and automates trace segmentation.
When the time between repetitions is not constant, further trace alignment techniques are required.
Building on \ac{vt}, we propose a simple method to learn representative segment templates from a profiling device similar to the victim, 
and to automatically locate and pull out these segments from other victim devices using simple pattern recognition.
We evaluate \ac{vt} on screaming channel attacks~\cite{camurati_UnderstandingScreamingChannels_2020}, 
which initially used a \acf{fc} known to appear at a single time in leaked signals, as a trigger to segment traces. 
We demonstrate that \ac{vt} not only performs equivalently to \ac{fc} on a standard attack scenario, but we also show how using \ac{vt} with the automatic pullout technique improves the attack efficiency and enables more realistic attack scenarios.
Thanks to \ac{vt}, screaming channel attacks can now: (1) succeed with only half of the segments collected compared to the \ac{fc} trigger from the original attack; and (2) absorb time variations between \acp{cp}.


\end{abstract}


\section{Introduction}
\acfp{sca}~\cite{standaert_IntroductionSideChannelAttacks_2010} exploit data-correlated leakages that arise when a device operates on data. The term \emph{side-channel} is used to denote physical leakage signals carrying confidential information. Side-channels are inherent to CMOS computing devices and can take many forms, from timing to power consumption to \ac{em} emanations. We call \emph{traces} to the sampled measurements of a side-channel during the execution of one or multiple targeted operations. The most common scenario in \acp{sca} targets a cryptographic key manipulated by \acfp{cp}, as eavesdropping such a key can jeopardize system confidentiality. We refer to these attacks as \acfp{scca}~\cite{choiTEMPESTComebackRealistic2020}. A trace can contain leakage from multiple \acp{cp}. Trace segmentation separates the different segments from the leakage, each corresponding to the side-channel measurement of one \ac{cp} execution.
During the attacking phase, a hypothesis $\hat{y} \in Y$ on the data $y$ is made ($Y$ = all possible data values). Techniques like 
\ac{dpa}\cite{kocher_DifferentialPowerAnalysis_}, \ac{cpa}\cite{brier_CorrelationPowerAnalysis_2004},  \ac{mia}\cite{gierlichsMutualInformationAnalysis2008a}, template attacks\cite{chari_TemplateAttacks_2003} and more recently \ac{dl}~\cite{masure_ComprehensiveStudyDeep_2019},
evaluate the relationship between $\hat{y}$ and the segment leakage values. Each hypothesis $Y$ is tested, and a probability is returned for all of them. The $\hat{y}$ having the highest probability is expected to correspond to the data $y$. 
For these attacks to work, it is important to know, for each segment point, to which \ac{cp} operations they belong. It makes it possible to find a relation between $y$ and leak values of the same \ac{cp} data-correlated operations present in all segments. To respect this requirement, segment synchronization is done during the collecting and pre-processing phases of the attack.


Focused on screaming channel attacks, we look into its collecting and pre-preprocessing phases, which aim at obtaining synchronized and denoised segments from the victim's leakage traces using techniques like time diversity (average multiple \acp{cp} computing the same data to reduce the noise).
To relax the triggering requirements for trace capturing and synchronization, our contributions include:

\begin{itemize}
    \item \emph{\acf{vt}}, a method to segment traces from side-channels of an embedded device executing a cryptographic software implementation. \ac{vt} requires neither external synchronization nor tampered victim software.
    \item  an experimental evaluation of the proposal on a realistic screaming channel attack to \ac{aes} on an embedded device.
    \item a discussion on the method limitations and solutions.
    \item experimental results demonstrating the gains obtained with these solutions.
\end{itemize}

\ac{vt} does not require any specific setup on the victim side, like a trigger signal to indicate the start of a \ac{cp}. 
It consists in finding a precise enough time duration of the targeted process executed periodically. This makes it possible to act as if a trigger would indicate a common location in all the process segments. This virtual trigger can then be used to segment \acp{cp} from a trace. 
The method aims at helping researchers to reduce the effort in target preparation and in the collecting phase of the attack, while also giving a small step toward a more realistic attack scenario.

The paper is organized as follows. Section~\ref{sec:Screaming channels} provides the context of this work and Section~\ref{sec:relatedwork} discusses related works. Then, Section~\ref{sec:method} details the proposed virtual trigger segmentation method, and Section~\ref{sec:experiments} evaluates it experimentally on a screaming channel setup. Section~\ref{sec:limitation} discusses the limitation of the method and proposes a solution to overcome it. Finally, Section~\ref{sec:conclusion} concludes the paper.

\section{Screaming channel attacks}
\label{sec:Screaming channels}
\begin{figure}[!t]
\centerline{\includegraphics[width=0.36\textwidth]{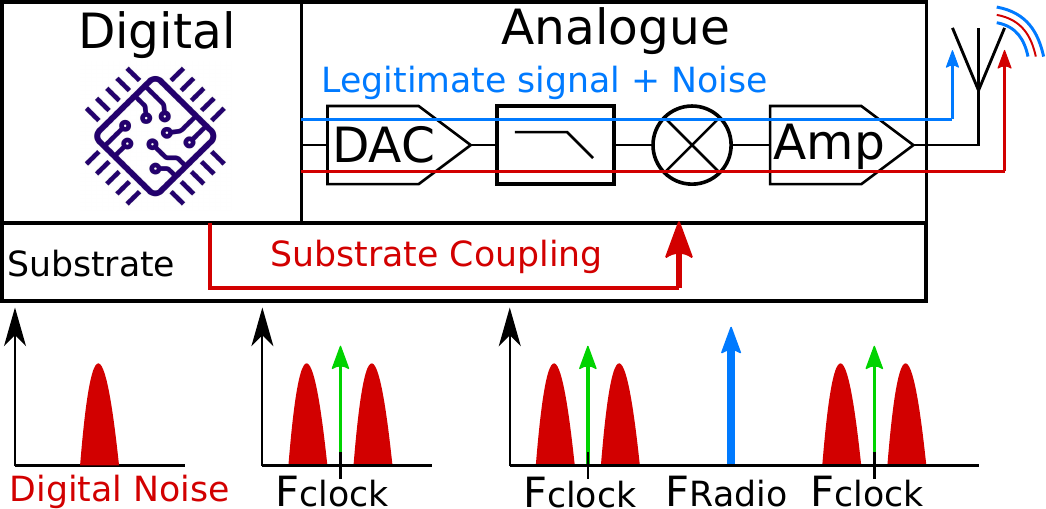}}
\caption{\textbf{Screaming channel attacks:} Conventional side-channels leak to the RF module in the analog part, present on the same die. This one transmits side-channels at a larger distance (until some meters).}
\label{Screaming}
\end{figure}

Experimental results build on the attack scenario called screaming channels introduced by Camurati, et al.~\cite{camurati_UnderstandingScreamingChannels_2020}. As illustrated in \figurename~\ref{Screaming}, screaming channels occur on mixed-signal devices where digital processing is collocated with analog \ac{rf} electronics over a single die. Side-channels originated from digital processing mix with \ac{rf} signal and get amplified, modulated, and broadcast. The primary threat posed by screaming channels is the risk of transmitting secrets over long distances, i.e., \emph{scream} them.

The screaming channel signals are very noisy. Plus, to collect them, it is necessary that the \ac{rf} module transmits a legitimate signal. In the context of this paper, it is a Bluetooth signal. Between two Bluetooth transmissions, the collected signal contains \emph{holes}. 
As in regular screaming channel analysis, to counterbalance these constraints, time diversity is used during the collection phase. 
The principle is to force the device to compute multiple encryptions with the same plaintext and key. Since the same data has been computed, their segment values should be the same, except for the noise. Averaging the segments returns a \ac{cp} segment with reduced noise.

\section{Related Works on segment synchronization}
\label{sec:relatedwork}
Synchronization is used to know to which \ac{cp} operations each segment point belongs to.
Otherwise, segment points corresponding to operations whose leakage values are data-correlated would be compared with other unrelated points. Therefore, it would be harder to distinguish a relationship between leakage values and data. We name these data-correlated points as \acfp{poi}.
To synchronize segments, an alignment between them can be done using techniques like static alignment~\cite{mangard_AttacksHiding_2007}, longest common sequence~\cite{jia_SideChannelLeakage_2020}, elastic alignment~\cite{abdellatif_EfficientAlignmentElectromagnetic_2019} and synchronous real-time sampling~\cite{yang_SynchronousRealTimeSampling_2021}. Before aligning segments, it is first necessary to locate these segments of interest in the traces. The most common technique in \ac{sca} research consists of inserting a trigger signal to start the trace measurement synchronized with the beginning of the \ac{cp}.
Attack setups are prepared to either have (1) the victim to create the trigger signal to inform the attacker when encryption starts, or (2) the attacker sending a trigger signal to the victim to make it start at a precise moment. This is an accepted scenario in the community to enable \ac{sca} research. But it assumes attackers have access to the victim to generate or listen to a trigger synchronization signal. SAKURA and NewAE's Chipwhisperer are widely used platforms in the community that follow this approach.

However, in many cases, using a trigger signal is impossible. For example, when a given firmware cannot be modified to add instructions that control the trigger signal. Or simply because the device used to collect traces is not capable of capturing two signals, the side-channel signal and the trigger signal, concurrently.  
Locating \acp{cp} in traces without using trigger signals can be done with pattern recognition techniques. For this purpose, Beckers, et al.~\cite{beckers_DesignImplementationWaveformMatching_2016} compare methods that calculate the correspondence between trace and pattern values. When this match score is over a pre-defined threshold, the corresponding part of the leakage is considered as being the location in the trace of a targeted segment. IcWaves\footnote{\url{https://www. riscure.com/security-tools/hardware/icwaves.}} implements such pattern recognition methods. 

Nevertheless, to use these methods, the attacker is supposed to already have a pattern or characterized segments having the same statistical properties as the researched segments, representative of the triggering moment. Therefore, the question of how to find this pattern remains open.
To that end, Trautmann et al.~\cite{trautmann_SemiAutomaticLocatingCryptographic_2022} proposed a technique to locate AES \acp{cp} in leakage signals by searching for parts of the leakage having consecutive similar patterns corresponding to the 10 AES rounds. This method can find AES CPs in long traces also containing other \ac{cp} operation leakages. Souissi et al.~\cite{souissi_NovelApplicationsWavelet_2011} used wavelet transforms to detect the limit of AES segments in traces and then used these segments to do pattern recognition. 

In the screaming channels context, in order to locate \acp{cp} from leakage signals, the only technique reported so far in the literature, by Camurati et al.~\cite{camurati_UnderstandingScreamingChannels_2020} and Wang et al.~\cite{wang_FarFieldEM_2020}, used a frequency component trigger mechanism\footnote{\url{https://github.com/bolek42/rsa-sdr.}}. The method continuously monitors a frequency band at which a given signal is present only at a unique instant of the targeted \acp{cp}. Segment locations should then correspond to trace locations where this signal is found.

In contrast with the proposed \ac{vt}, all these methods require specific equipment other than the ones needed for the \ac{sca} itself, like a spectrum analyzer~\cite{camurati_UnderstandingScreamingChannels_2020,wang_FarFieldEM_2020} or a powerful GPU~\cite{trautmann_SemiAutomaticLocatingCryptographic_2022}. 
Therefore, before using more complex, expensive, or time-consuming solutions, \ac{vt} is a trace segmentation method that can be used when the targeted device runs \acp{cp} without interruptions, in order to reduce the collecting phase complexity.

\section{Considered attack scenario}
\label{sec:attack_scenario}
We consider a passive attack on system confidentiality 
exploiting power or \ac{em} side-channels from a \ac{cp} manipulating \ac{aes} keys. 
This causes leakage propagation through side-channels with very low \ac{snr}. 
The attacker capabilities include: they (1) know the precise \ac{cp} duration 
(or can use the method presented later in the paper to find it)
and (2) have access to a copy of the victim device used to build a profiled attack.

When collecting segments of interest, either during the profiling or the attacking phase, the following steps retrieve $N$ segments from a collected trace and average them to generate a final segment with reduced noise:
\begin{enumerate}
    \item The victim starts running a series of encryptions using the same plaintext and key.
    \item The attacker, using a \ac{sdr}, starts collecting the trace after a certain delay to make sure \ac{cp} executions have begun. 
    \item The collected trace is cut into $N$ segments. To make sure segmentation happens only while \acp{cp} are present, $N$ is chosen in a way that \emph{$N\times$\ac{cp} duration} is inferior to the \acp{cp} series execution duration. Except for the noise, these segments carry the same information, as each \ac{cp} computed the same data.
    \item A fine alignment is done between these segments, and the $N$ aligned segments are averaged together, yielding a single representative \ac{cp} segment with reduced noise (see Algorithm in \figurename~\ref{alg1}).
\end{enumerate}

\begin{figure}[!t]
\begin{algorithmic}[1]
\small
    \Require Misaligned \ac{cp} segments: $Segs_{in}$ \Comment{Segments misaligned}
    \Ensure Aligned \ac{cp} segment: $Seg_{out}$ \Comment{Segment aligned}
    \State $template  \leftarrow \text{None}$ 
    \For{$index~IN~len(Segs_{in})$}
        \If{$\text{template == None}$}
            \State $template \leftarrow Segs_{in}[index]$
        \Else
            \State $shift \leftarrow max(corr(Segs_{in}[index],~template) )$
            \State $Segs_{in} \leftarrow ShiftValues(Segs_{in},~shift)$
            \State $template \leftarrow average(Segs_{in}[:index],~axis=0)$
        \EndIf
    \EndFor
    \State $Seg_{out} \leftarrow average(Segs_{in},~axis=0)$

\end{algorithmic}
\caption{\textbf{Algorithm:} Fine alignment between segments}
\label{alg1}
\end{figure}

\section{Proposed Virtual triggering method}
\label{sec:method}
\begin{figure}[!t]
\centerline{\includegraphics[width=0.5\textwidth]{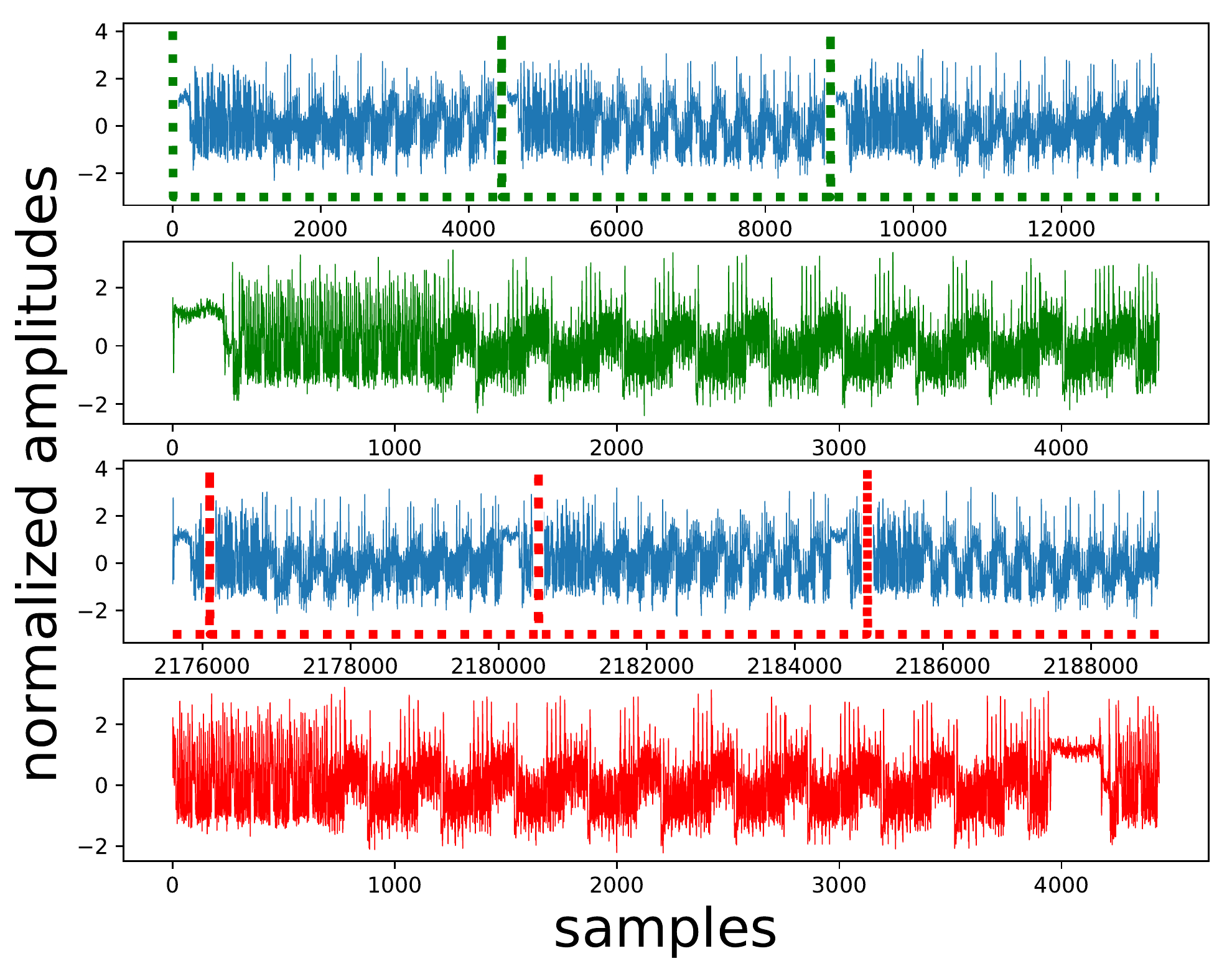}}
\caption{\textbf{Segment \acp{cp} using the virtual trigger:} \textit{1st row:} the beginning of the collected trace and the \ac{vt} positions determined according to the known \ac{cp} length. \textit{2nd row:} a segment aligned to segment start time. \textit{3rd/4th rows:} if the length is not precise enough, a drift appears in start times, and hence \ac{vt} indicates a very different point in \acp{cp} at the end of the trace.}
\label{Cut_COs_segment}
\end{figure}

\acf{vt} aims at cutting each captured trace in segments (step 3 of trace collection, Section~\ref{sec:attack_scenario}), with a distance between cuts being closely enough to the \ac{cp} length $L_{cp}$ to correctly segment \acp{cp} in the trace.
If $L_{cp}$ is not known very precisely, as illustrated in \figurename~\ref{Cut_COs_segment}, an offset appears in the \ac{vt} position.
Consequently, segment starting points would be triggered at different instants in the \acp{cp}, making the final segments too different from each other to be aligned.
This would deteriorate the denoised \ac{cp} segment, as averaged samples would not correspond to the same instructions. 

\subsection{Find accurate \ac{cp} length $L_{cp}$}
We propose a two-step method to find the accurate \ac{cp} length. This step can be performed only one time as $L_{cp}$ is the same for all \ac{cp} segments. First, an approximate value $\hat{L}_{cp}$ is found, reducing the effort to find the accurate length $L_{cp}$ in the second step.

The first step consists in exploiting one trace containing multiple \acp{cp}. Auto-correlating such a trace, a peak is obtained each time \acp{cp} are aligned together.
Therefore, $\hat{L}_{cp}$ can be approximated by manually measuring the average distance between two peaks.

The second step aims at finding the precise length $L_{cp}$ of one \ac{cp}, which is equal to the approximate length plus one $\delta$. 
The method used to find the value of this $\delta$ is presented in Algorithm in~\figurename~\ref{alg2}. It consists of testing a range $\Delta$ of $\hat\delta$ candidates. It segments the trace with the corresponding length: $\hat{L}_{cp}$ + $\hat{\delta}$ (line 5), separates these segments into two equal groups, to distribute at best one group contains the even segments, the other the odd segments, averages each group to get two different reference segments A and B (line 6~-~7). 
Then, the L1 distance (line 8) between the two reference segments should be at its minimum when the most accurate $\hat\delta$ value is used. This distance is computed using equation \eqref{eq_Diff}.

\begin{equation}
\textit{L1 Distance} = \frac{\textit{abs(Segment A - Segment B)}}{\textit{NB samples in one segment}}\label{eq_Diff}
\end{equation}

where $abs$ is the absolute value, and \emph{NB samples in one segment} is the number of samples in one segment.

\begin{figure}[!t]

\begin{algorithmic}[1]
\small
    \Require Approximate \ac{cp} length: $\hat{L}_{cp}$
    \Require Range of $\delta~candidate~values: I_{nterval}$
    \Ensure Accurate \ac{cp} length: $L_{cp}$ 
    \State $\hat\delta \leftarrow - I_{nterval} / 2 $ \Comment{Delta}
    \State $S  \leftarrow I_{nterval} / 2 $ \Comment{Stop}
    \State $\delta_{step} \leftarrow I_{nterval} / \text{number of steps} $ \Comment{Delta step}
    \Repeat 
        \State $Segs \leftarrow \text{Cut the trace in N segments} \text{  of length = $\hat{L}_{cp} + \hat\delta$ }$
        \State $Segment_A \leftarrow average(pair~Segs)$
        \State $Segment_B \leftarrow average(odd~Segs)$
        \State $Distance(\hat\delta)\leftarrow Segment_A - Segment_B$
        \State $\hat\delta \leftarrow \hat\delta + \delta_{step} $
    \Until {$\text{$\hat\delta$} > \text{S}$}
    \State $L_{cp} \leftarrow \hat{L}_{cp} + \hat\delta~\text{for which}~Distance(\hat\delta) \text{ has its lowest value}$

\end{algorithmic}
\caption{\textbf{Algorithm:} How to find accurate \ac{cp} length $L_{cp}$}
\label{alg2}
\end{figure}

\subsection{Align the segments together}

After having found the accurate length $L_{cp}$ of a single \ac{cp} execution, it is possible to segment correctly any trace containing the same \acp{cp} executed recursively. 
All the obtained denoised segments contain leakage samples from the same instructions but are not aligned together since segmentation begins at a random point in each trace.
To have aligned segments, it is important to make them begin at a common point in the \ac{cp}.
For that, it is possible to take one of the segments as a reference and shift the values of the others until they reach the position where they best correspond to the reference segment. 
Afterward, a fine alignment can get them perfectly aligned to each other.

\section{Experiments}
\label{sec:experiments}
\subsection{Experimental setup}
\begin{figure}[!t]
\centerline{\includegraphics[width=0.225\textwidth]{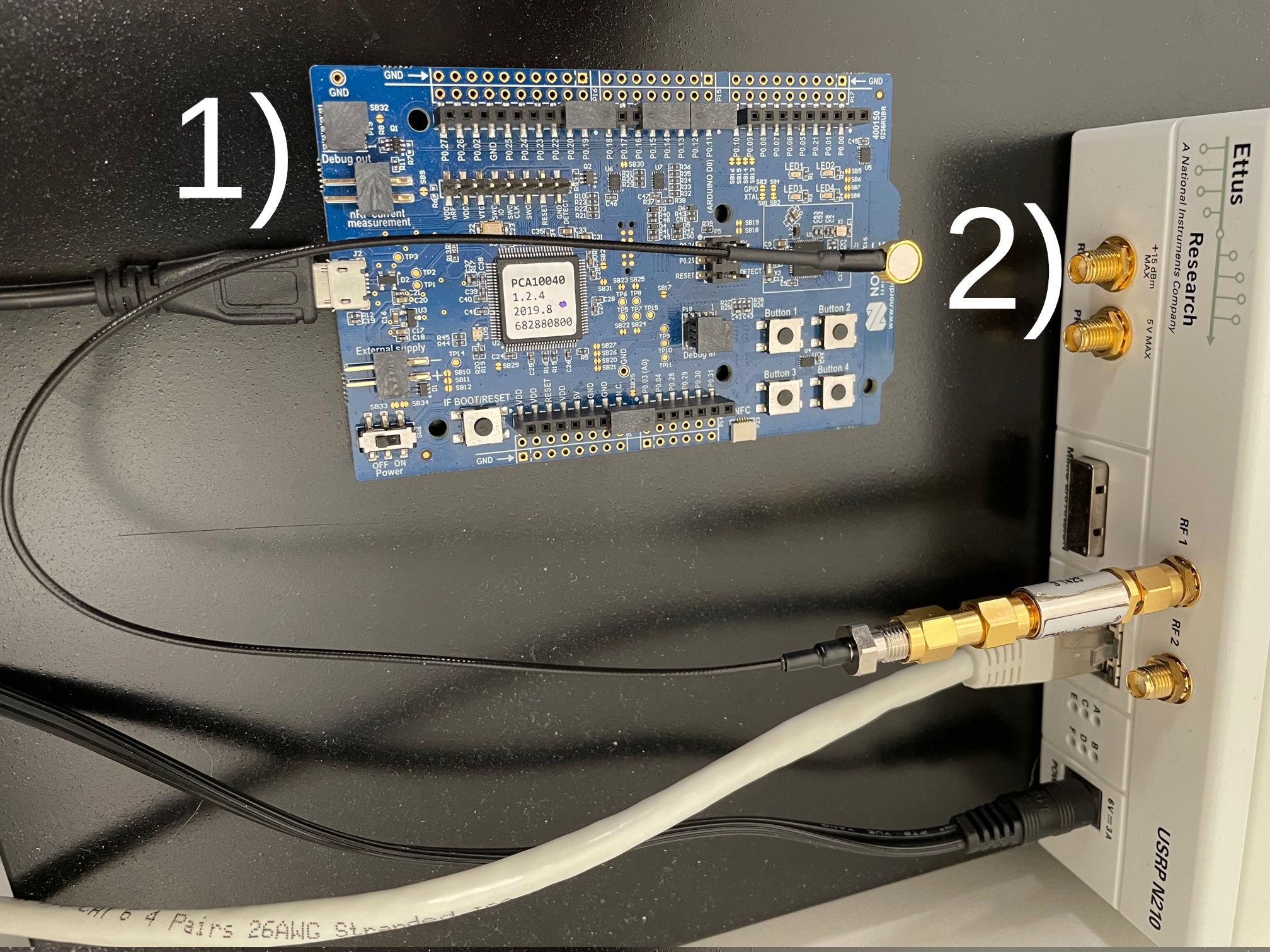}}
\caption{\textbf{The setup}: 1) The victim PCA10040 device running AES encoding and transmitting a Bluetooth signal at 2.4GHz. 2) The SDR device collects the victim's data-correlated leakage at 2.528GHz.}
\label{setup}
\end{figure}

Fig.~\ref{setup} shows the setup.
It comprises a victim board, a mixed-signal Nordic Semiconductor chip as in the seminal screaming channels work~\cite{camurati_UnderstandingScreamingChannels_2020} (PCA10040\footnote{\url{https://www.nordicsemi.com/Software-and-tools/Development-Kits/nRF52-DK.}}), processing software-based AES encryptions. 
On the attacker side, an \ac{sdr} captures the victim RF emissions: a USRP N210\footnote{\url{https://www.ettus.com/all-products/un210-kit/}} with an SBX daughterboard covering a frequency band from 400MHz to 4.4GHz. Instructions are sent to the victim to initialize the new plaintext value and launch a series of $N=500$ encryptions. The signal captured by the SDR is received by the attacker's computer, which runs the pre-processing and decoding process. 

\subsection{Finding the accurate \ac{cp} length $L_{cp}$}
\begin{figure}[!t]\centerline{\includegraphics[width=0.45\textwidth]{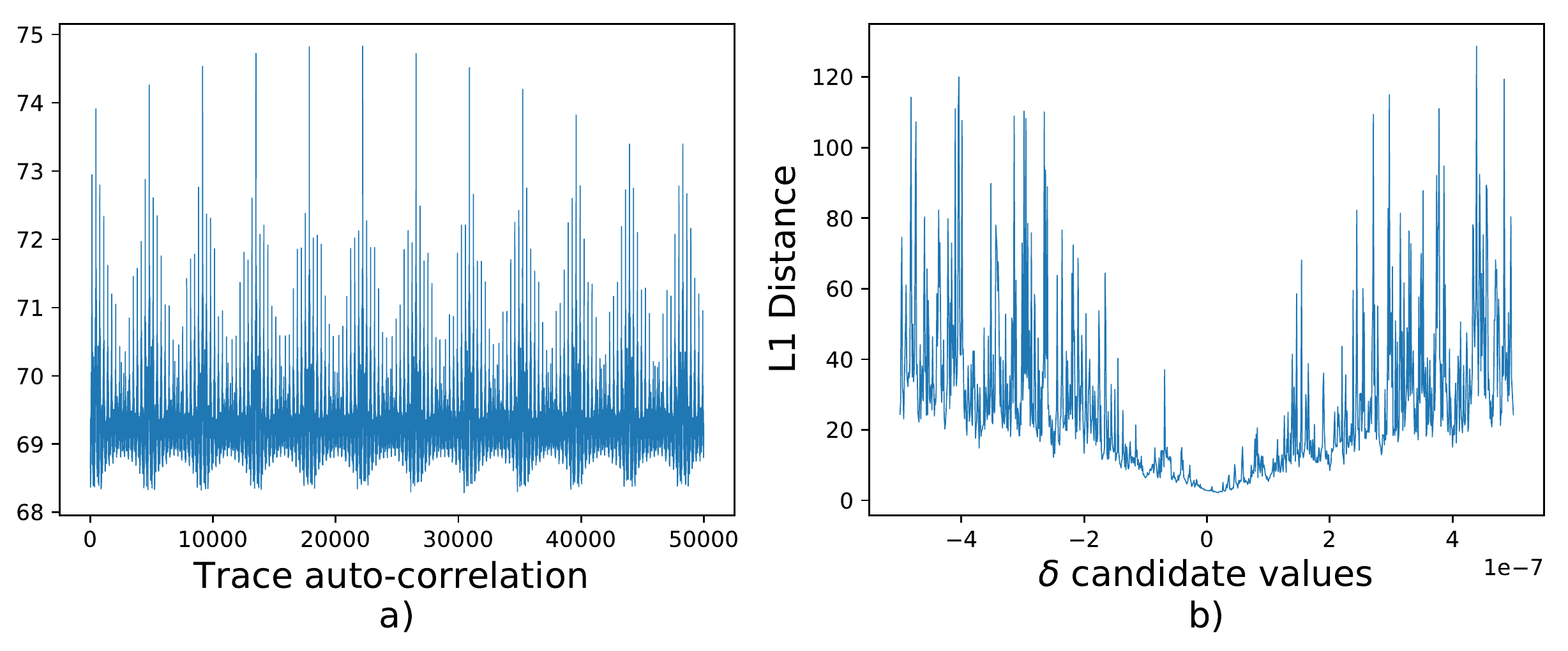}}
\caption{\textbf{Find \ac{cp} segment length}: a) Approximate length: trace auto-correlation returns a series of peaks. Each peak occurs when \acp{cp} are aligned. Manual measurement gives an approximate \ac{cp} length. b) Precise length: the closer the $\hat\delta$ candidate value is to the real value, the more similar the two segments A and B are. Here the best $\hat\delta$ candidate (min. L1 distance) is 18ns.}
\label{CP_length}
\end{figure}

The auto-correlation results after applying the first step to find the approximate \ac{cp} segment length are shown in Fig.~\ref{CP_length}~a). The average distance between two correlation peaks gives an approximate length of 4350 samples, which considering a sample rate of 5MHz, gives an approximate \ac{cp} length of 870us. 

The second step is performed to find a more accurate \ac{cp} length. The algorithm in~\figurename~\ref{alg2} runs on a range of 1000 ns (from -500 ns to 500 ns) with a step of 1ns. When the difference between the two \ac{cp} appears to be at its minimum, this means $\hat\delta$ is close to the real $\delta$ value. As shown on \figurename~\ref{CP_length}~b), the best value is found at 18ns. Automating this step is kept for future work. It would consist in automatically detecting the $\hat\delta$ value having the lowest result. For that, it would be possible first to detect the range containing the lower minima.

\subsection{Segmentation and alignment}
\begin{figure}[!t]
\centerline{\includegraphics[width=0.45\textwidth]{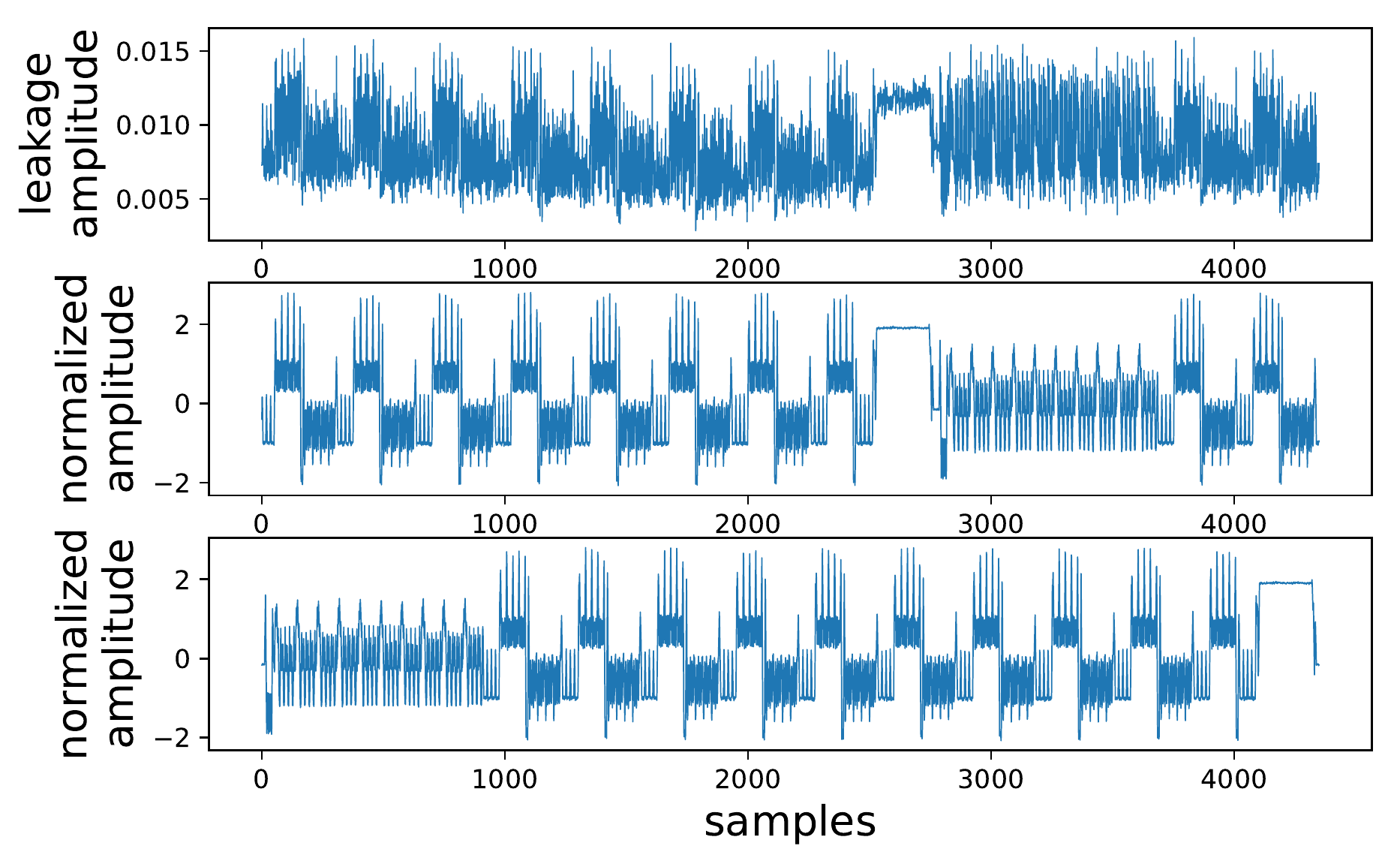}}
\caption{\textbf{Obtain denoised and align \ac{cp} segments:} \textit{1st row:} shows a segment in the initial collected trace. \textit{2nd row:} corresponds to averaged segments from trace cut with a precise length. \textit{3rd row:} is the average segment aligned to a recognizable point, to be aligned with all other denoised segments. The ten rounds of AES are visible.}
\label{CP_SCA}
\end{figure}

Knowing the precise \ac{cp} length, the trace is segmented (first row of Fig.~\ref{CP_SCA} shows one segment), and the segments are finely aligned together and averaged (second row of Fig.~\ref{CP_SCA}). 
In the denoised segment, a specific instant of \acp{cp} is localized, and the segment values are shifted to make the segment begin at this instant (third row of Fig.~\ref{CP_SCA}). 
In this case, the time between two encryptions is easily recognizable as it does not vary. By computing the variance with a window of its size (230 samples), we can detect it as this variance has its lowest result at the same \ac{cp} location in all segments.
If the segment starting point is placed in the middle of this instant, it is then undetectable since the 230 samples are cut into two parts, one at the beginning of the segment, and the second at the end. To anticipate these cases, the operation is performed on the denoised segment concatenated with its own 230 first samples. Therefore, this instant can be reconstituted at the end of the segment and be detectable.

With this final step, a coarse-grained alignment is done between all denoised segments. A fine alignment (Algorithm in \figurename~\ref{alg1}) can be necessary to have them even better aligned.

\subsection{Comparison with original screaming channel attack}
To evaluate \ac{vt}, different experiments are run to compare against the original method~\cite{camurati_UnderstandingScreamingChannels_2020}. 
Denoised \ac{cp} segments are extracted from the same collected traces using both the \ac{vt} and the frequency component triggers. 5000 segments are used to build a profile, and 200 for the attack phase. We evaluate attack efficiency with the \acf{pge} of each AES key byte. The \ac{pge} corresponds to the rank of the correct key byte value in a list of the possible hypothesis on data from the decoding process.

\begin{figure}[!t]
\centerline{\includegraphics[width=0.4\textwidth]{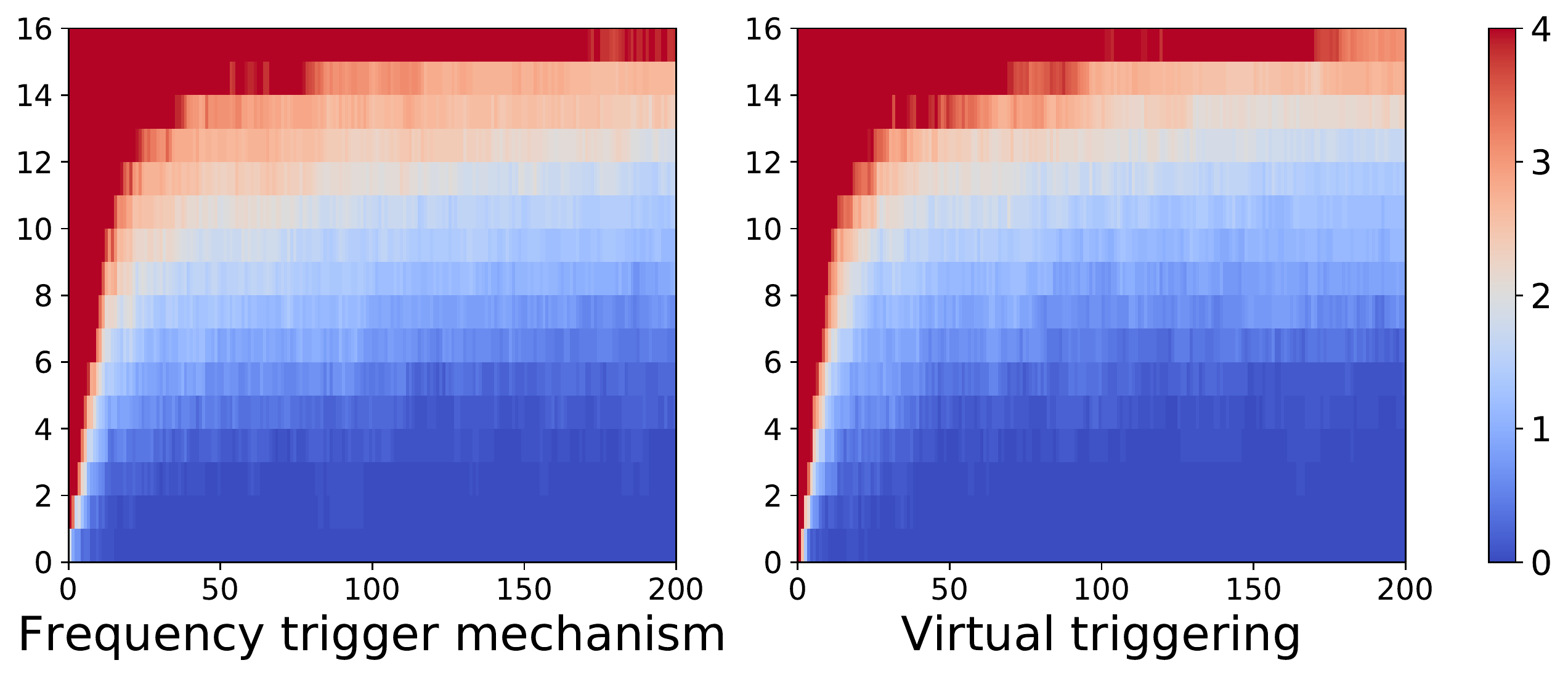}}
\caption{\textbf{Comparison between \ac{fc} and \ac{vt} triggers} as \ac{pge} per key byte (the higher the blue array, the better, 16 is a maximum): the proposed method does not decrease the attack efficiency for screaming channel attacks while reducing the segmentation complexity.}
\label{C_vs_Mth}
\end{figure}

Fig.~\ref{C_vs_Mth} plots the average \ac{pge} over 30 attacks of each of the 16 key bytes sorted from best to worst, versus the number of traces used in the attack phase.
The darker the blue, the lower the \ac{pge} is, i.e., the closer the attack is to guess the correct key byte. If \ac{pge} is equal to 0, the hypothesis made on the key byte is correct; if it is equal to $n$, then $n$ hypothesis have a higher probability than the correct one. The red color corresponds to a \ac{pge} higher or equal to 4,
which is the \ac{pge} value that no more than one key byte should reach for the key being easily recoverable in a few seconds using brute force.
We can see that the \acp{pge} are equivalent for both methods: 100 traces are enough in most cases to have only one byte with a \ac{pge} superior or equal to 4. 
These results show that \ac{vt}, a simpler and less expensive method, achieves an equivalent efficiency to the original attack~\cite{camurati_UnderstandingScreamingChannels_2020}.

\begin{figure}[!t]
\centerline{\includegraphics[width=0.4\textwidth]{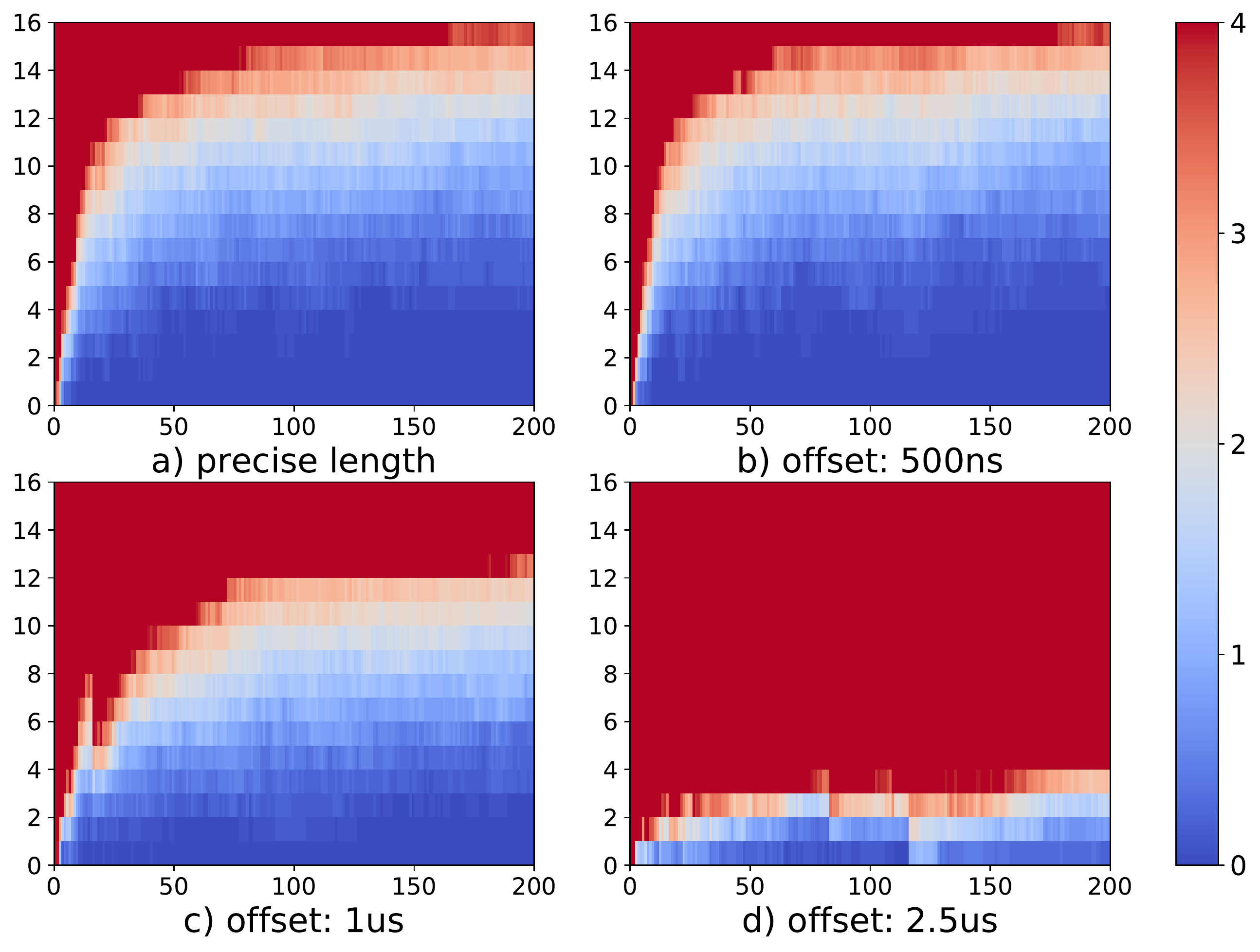}}
\caption{\textbf{Attack efficiency for different \ac{cp} length precisions}: The top left is the collection with a known length and $1 ps$ precision. Other results are collected with an offset of respectively $500ns$, $1\mu s$ and $2.5\mu s$.}
\label{precision}
\end{figure}

We also evaluate \ac{vt} according to the \ac{cp} length precision. The top left result in \figurename~\ref{precision} shows the attack result after a segmentation with the most precise length we were able to determine (precision: $1ps$). The attack was repeated forcing an offset of respectively $500ns$, $1\mu s$, and $2.5\mu s$. It can be seen that an offset of $500ns$ does not deteriorate segment information, but from $1\mu s$ the attack efficiency decreases.

\subsection{Generalization of the method}
\begin{figure}[!t]
\centerline{\includegraphics[width=0.4\textwidth]{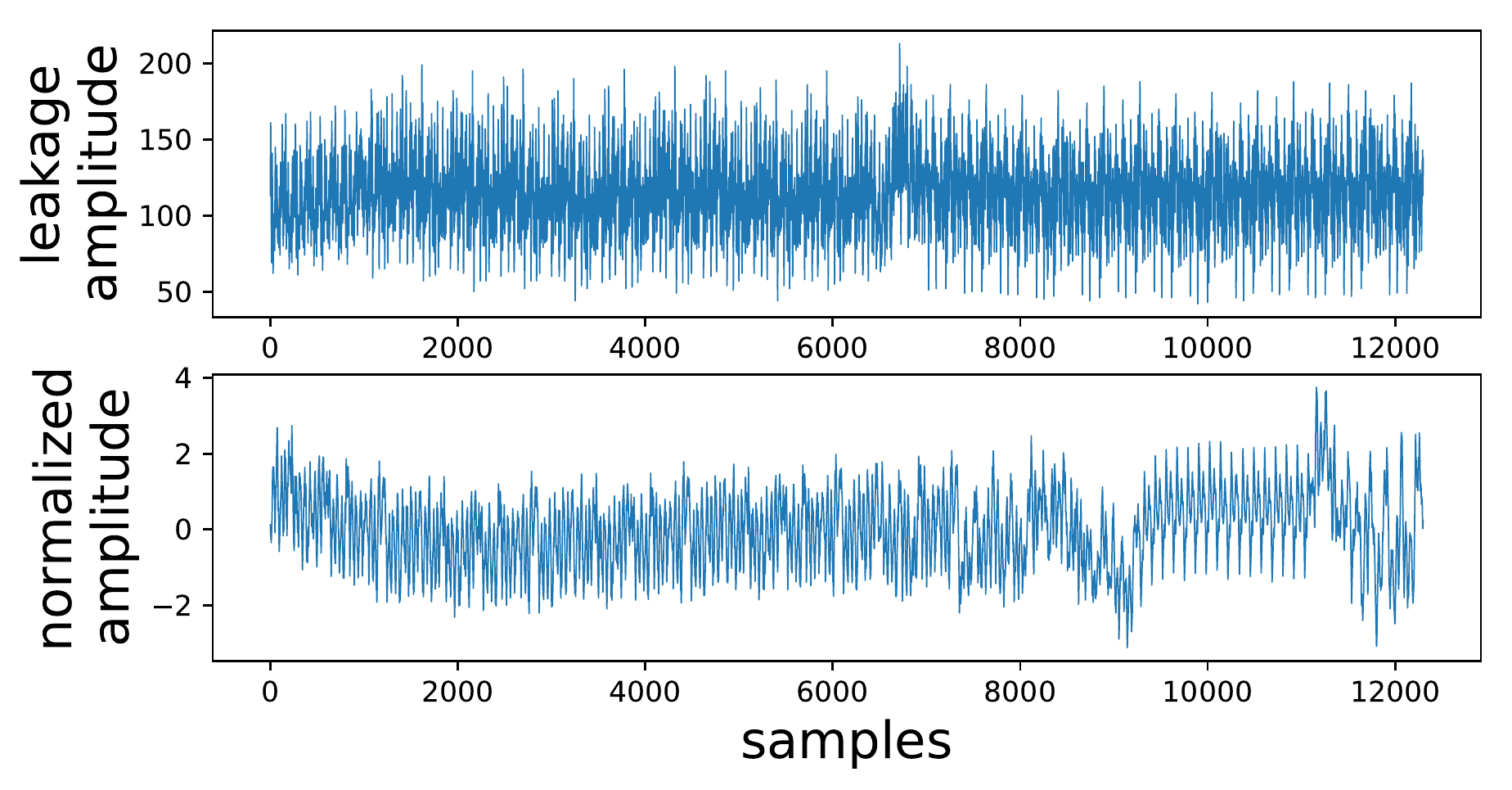}}
\caption{\textbf{Virual trigger applied for segmentation of 500 AES of a secure boot process:}. First row: initial trace. Second row: Denoised Segment.}
\label{CP_boot}
\end{figure}

To test the generalization of \ac{vt} beyond the screaming channels context, hence on regular power/EM traces, we apply it to a signal collected by other authors~\cite{trautmann_SemiAutomaticLocatingCryptographic_2022}, using leakage from a MCU device (STM32F303) executing a secure boot process containing 500 AES encryptions. 
As one of the requirements of our method is to have a signal containing only a series of \acp{cp}, we kept only the part of their signal containing AES segments.
Using \ac{vt}, the denoised segment in Fig.~\ref{CP_boot} is obtained.

\section{Improving \ac{vt} with pattern recognition}
\label{sec:limitation}
\ac{vt} enables pinpointing reliable points from \acp{cp} by precisely computing segments length, hence acting as if a trigger signal was used.
In turn, this makes it possible to segment traces in a way that all final segments are aligned with each other.
However, when delays vary between the starting instants of segments, or if the attacker targets a device executing \acp{cp} asynchronously, the method does not hold.

In these contexts, \ac{vt} can still be used during a profiling phase to build a template of the \ac{cp} segments for later use during an attack. Next, we propose a simple technique that builds on \ac{vt} to: (1) locate and extract a denoised \ac{cp} segment from a profiling device (a copy of the victim), and learn a \emph{segment template}; (2) automatically locate and \emph{pullout} \ac{cp} segments from traces collected on a victim instance of the device. We also demonstrate how, interestingly, this technique can absorb arbitrary time variations between \acp{cp} and improve the efficiency of existing attack methods.

\emph{During profiling}, the attacker can force the device, e.g., with resets, to repeatedly execute a program containing \acp{cp} without time variations. 
Afterward, \emph{during the attack phase}, the learned segment template (or only a subpart of it corresponding to a \ac{cp}) can be used to locate \ac{cp} segments in the traces using pattern recognition techniques.
\emph{For pattern recognition}, we cross-correlate traces with the learned pattern and detect peak positions indicating \ac{cp} locations.

\begin{figure}[!t]
\centerline{\includegraphics[width=0.4\textwidth]{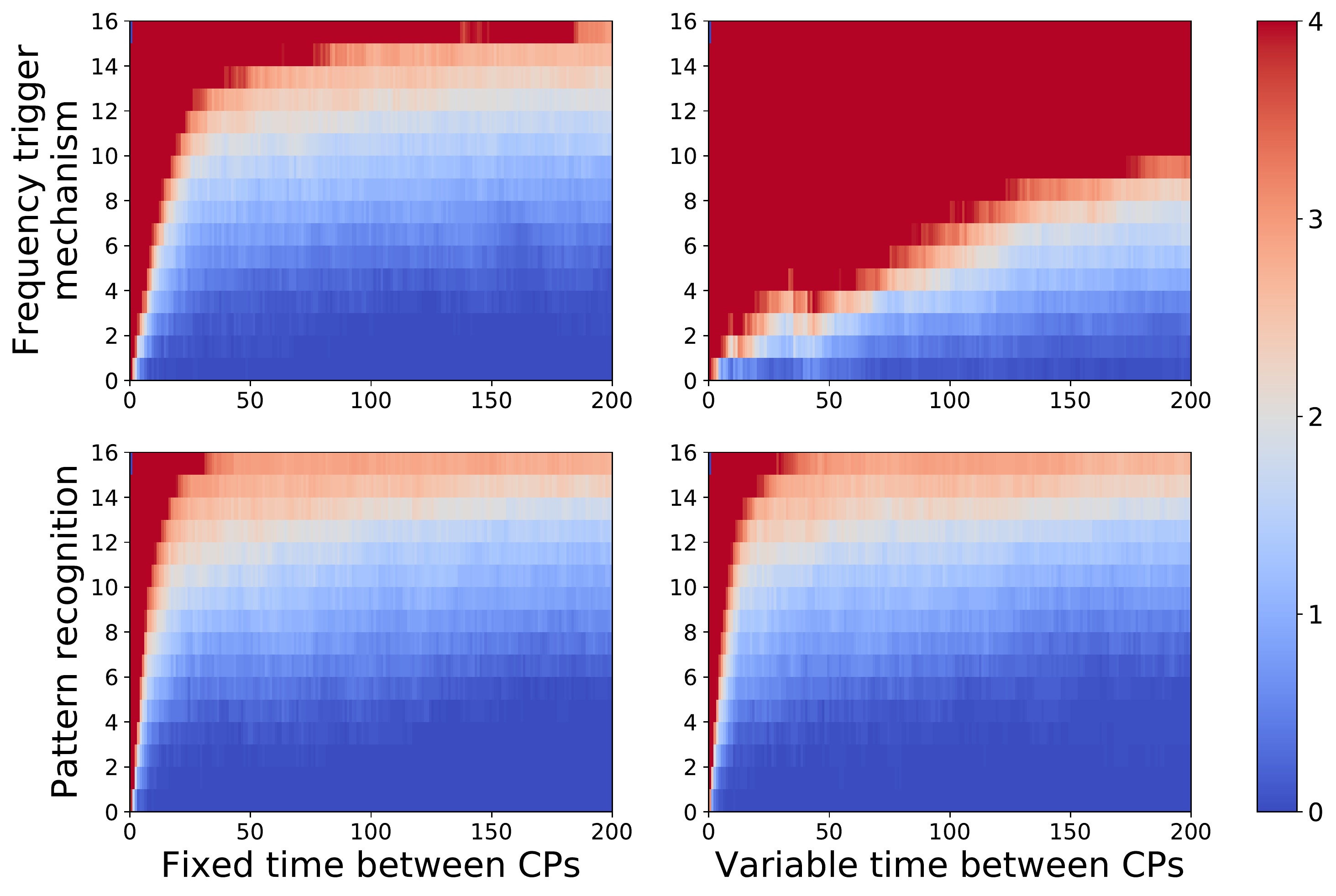}}
\caption{\textbf{Comparison between \ac{fc} trigger and pattern recognition:} Top figures show that the original screaming channels method does not resist to time variations between \acp{cp}. Bottom figures show how \ac{vt} and automatic pattern pullout is resistant to time variations between \acp{cp}.}
\label{PR_vs_C}
\end{figure}

We compare our results with the efficiency of the \ac{fc} trigger. 
The experiment has been performed in two scenarios: one using periodic encryptions and another with a variable delay between them. \figurename~\ref{PR_vs_C} shows how using pattern recognition significantly increases the attack efficiency. As previously with \ac{fc} trigger, around 100 segments are needed. Using pattern recognition with the \ac{vt} extracted pattern, less than 50 segments are enough.

An obvious advantage of \ac{vt} compared to the pattern pullout technique with pattern recognition is that \ac{vt} needs less computation effort, hence the collection phase is faster. In our case, collecting 10K denoised segments took 20 hours using pattern recognition, when only 5 hours were needed with \ac{vt}.
However, using the pattern pullout technique enables attacks in scenarios with time variations between \acp{cp}, which can come from system interrupts or simple countermeasures using time randomization. Nonetheless, we have not verified the method in these conditions, this is left for future work.

\section{Conclusion}
\label{sec:discussion}\label{sec:conclusion}
This paper has proposed \ac{vt}, a simple-to-implement method to collect denoised \ac{cp} segments, exploiting time diversity in the context of screaming channels. The efficiency of \ac{vt} is demonstrated with respect to the state-of-the-art screaming channel trace collection method~\cite{camurati_UnderstandingScreamingChannels_2020}.
Compared to more complex and recent methods to find \acp{cp} in traces containing other operations than those exploited by the attack~\cite{trautmann_SemiAutomaticLocatingCryptographic_2022}, \ac{vt} is much simpler.
While our experimental results concentrate on screaming channels, \ac{vt} and the pattern pullout technique are generic and can be used for other forms of physical leakages.

\bibliographystyle{ieeetr}
\bibliography{biblio}

\end{document}